\documentclass{cpcauth}

\usepackage[numbers,square,comma]{natbib}
\usepackage{epsfig}

\newcommand{\be}{\begin{equation}}
\newcommand{\ee}{\end{equation}}
\newcommand{\sign}{\mathop{\mathrm{sign}}\nolimits}

\begin{document}
%%%%%%%%%%%%%%%%%%%%%%%%%%%%%%%%%%%%%%%%%%%%%%%%%%%%%%%%%%%%%%%%%%%%%%%%%%%

%%%%%%%%%%%%%%%%%%%%%%%%%%%%%%%%%%%%%%%%%%%%%%%%%%%%%%%%%%%%%%%%%%%%%%%%%%%
% TITLEPAGE
%%%%%%%%%%%%%%%%%%%%%%%%%%%%%%%%%%%%%%%%%%%%%%%%%%%%%%%%%%%%%%%%%%%%%%%%%%%
\begin{frontmatter}
%\date{\today}

\title{
%\vspace{-5.0cm}
%\begin{flushright}
%{\normalsize UNIGRAZ-}\\
%{\normalsize UTP-dd-dd-yy}\\
%\end{flushright}
%\vspace*{2.5cm}
Improving the Dirac Operator in Lattice QCD}
\author[label1]{Christof Gattringer\thanksref{label3}},
\thanks[label3]{Supported by the Austrian Academy of Sciences (Apart
654)}
\author[label2]{C.~B. Lang}
\ead{christian.lang@uni-graz.at}
\address[label1]{Inst. f. Theoret. Physik, Univ. Regensburg, 
Universit\"atsstr. 31,  D-93053~Regensburg,  Germany}
\address[label2]{Inst. f. Theoret. Physik, Universit\"at Graz, A-8010~Graz, AUSTRIA}

\begin{abstract}
Recently various new concepts for the construction of Dirac operators in
lattice Quantum Chromodynamics (QCD) have been introduced. These operators
satisfy  the so-called Ginsparg-Wilson condition (GWC), thus obeying the
Atiyah-Singer index theorem and violating chiral symmetry only in a modest and
local form. Here we present studies in 4-d for SU(3) gauge  configurations with
non-trivial topological content. We study the flow of eigenvalues  and we
compare the numerical stability and efficiency of a recently suggested chirally
improved operator with that of others in this respect.  
\end{abstract}

\end{frontmatter}

%%%%%%%%%%%%%%%%%%%%%%%%%%%%%%%%%%%%%%%%%%%%%%%%%%%%%%%%%%%%%%%%%%%%%%%%%%%
% MAIN PART OF PAPER
%%%%%%%%%%%%%%%%%%%%%%%%%%%%%%%%%%%%%%%%%%%%%%%%%%%%%%%%%%%%%%%%%%%%%%%%%%%

\section{Motivation and technicalities}

A priori continuum Dirac operators have chiral symmetry, which then  is broken
spontaneously in the full QCD dynamics. Lattice discretizations of the Dirac
operator are bound to violate the chiral symmetry. However, if they obey the
GWC this violation is in its mildest form. Currently three types of exact
realizations of so-called Ginsparg-Wilson (GW) operators are known:  The
overlap operator, perfect actions and domain wall fermions; for reviews on
recent developments cf. \cite{reviews}. In  \cite{Ga00} we suggested a
method to systematically expand the lattice Dirac operator in terms of a series
of simple lattice operators  and to turn the GWC into a large algebraic system
of coupled equations for the expansion coefficients. The solution for a finite
parametrization leads to an approximate GW-operator. Such a chirally improved
Dirac operator has been constructed in 2-d for the Schwinger model
\cite{GaHi00} and also for 4-d QCD \cite{GaHiLa00},  
where an analysis of its spectrum revealed intriguing
properties.

The spectral properties of QCD Dirac operators provide an efficient means  to
study properties of the  QCD vacuum related to chiral symmetry breaking. The
spectral density in the limit of vanishing eigenvalues is proportional to the
order parameter of chiral symmetry breaking via the Banks-Casher relation
\cite{BaCa80}. The microscopic distribution density exhibits features in a
universality class discussed in random matrix theory \cite{VeWe00}. The
eigenmodes for exactly vanishing eigenvalues appear to represent instantons
(classical solutions of the QCD field equations). In the quantization of a
lattice-regularized field theory it may come as a surprise that instanton-like
objects (originally defined on differentiable manifolds) may play a role.
Indeed, they are a miniscule contribution to the total action of a gauge
configuration. However, as will be seen, they are observed in realistic gauge
field configurations and they characterize the sector of small eigenmodes
\cite{DeHaEdHeBlChCrHoIsMc,GaGoRa}.

The overlap operator \cite{NaNe} has the form
\begin{equation}\label{ovdef}
D_{ov} = 1 - Z\quad \textrm{with}\quad Z\equiv\gamma_5\,\sign(H)\;,
\end{equation}
where $H$ is related to the hermitian Dirac operator. This is
constructed from  an arbitrary Dirac operator $D_0$
(like e.g. the Wilson operator), $H=\gamma_5 \,(s-D_0)$,
and $s$ is a parameter which may be adjusted in order to minimize the
probability for zero modes of $H$. 
In case $D_0$ is already an overlap operator one reproduces 
$D_{ov}=D_0$ for $s=1$ due to 
$\sign\left(\sign(H)\right)=\sign(H)$.

The sign-function may be defined through the spectral representation,
but this is technically impossible for realistic QCD Dirac operators, which for
$L^4$ lattices have dimension $\mathcal{O}(10^5-10^6)$  ($\sim n_{color}\cdot
n_{Dirac} \cdot L^4 $). Note, that in the  subsequent applications the operator
has to be  determined many times, since it may be entering a diagonalization or
a  conjugate gradient inversion tool. One therefore relies on the relation
$\sign(H)=H/\sqrt{H^2}$ and approximates the inverse square by some method. In
our computations \cite{GaGoLa01} we follow the methods discussed by
\cite{HeJaLu99,HeJaLe00b,Bu98}.  
One approximates the inverse square root by a Chebychev
polynomial, which has exponential convergence in $[\epsilon,1]$, where
$\epsilon$ (and thus the order of the polynomial)  depends on the ratio of
smallest to largest eigenvalue of  $H^2$. Clenshaw's recurrence formula
provides further numerical stability.

Depending on the input Dirac operator $D_0$ the rate of convergence may be
quite unfavorable due to small eigenvalues of $H^2$.  Therefore one computes
the inverse square separately for the subspace of the smallest 20 eigenvalues
(using the spectral representation) and the reduced operator (with polynomial
approximation).  The subspace has to be determined with high accuracy.  All
diagonalizations have been done with the Arnoldi method
\cite{Arnoldi}.

\section{Flow of eigenvalues}
\label{Flow}

Lattice Dirac operators are necessarily non-hermitian, but 
``$\gamma_5$-hermitian'', i.e. one may always define a ``hermitian Dirac
operator'' $\gamma_5\,D$. The eigenvalues of a realistic ``hermitian (e.g.
Wilson-) Dirac operator'' are distributed broadly on the real axis and
have also values close to the origin. The sign-operator in (\ref{ovdef})
projects all $n_+$ positive eigenvalues to $+1$, and  all $n_-$ negative
eigenvalues to $-1$. The difference $n_+ - n_- $ is twice the number of zero modes
of $D_{ov}$. 

We are interested in the case when $(1-D_0)$ is close to, but not quite an
overlap operator. In particular we want to study here the emergence or
vanishing of zero modes. As a test case we construct a random, hermitian matrix
operator like $H$ with eigenvalues $\pm 1$. For $\gamma_5$ we choose an
arbitrary realization like $\Gamma_{ij}=(-1)^i\,\delta_{ij}$. The eigenvalues
of the corresponding $Z$ are then distributed randomly on a unit circle around
the origin. Due to ``$\gamma_5$-hermiticity'' of $Z$ its eigenvalues occur in
complex conjugate pairs. Real eigenvalues are at positions $\pm 1$ giving rise
to vanishing eigenvalues of $D_{ov}$, hence we call them zero modes.

\begin{figure}[bt]
\begin{center}
\epsfig{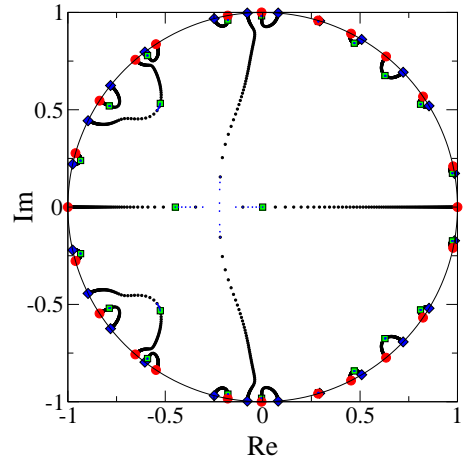}
\end{center}
\caption{\label{EVOnTheMove}
We display the flow of eigenvalues of a matrix $Z$ derived from a randomly constructed
matrix $H$ with all eigenvalues $\pm 1$ except for one, which
is slowly changed between $+1$  and $-1$  (begin and end positions:
full circles, value  0: square, intermediate values: dots).}
\end{figure}

Starting with a matrix where $n_+= n_-$ we will find no such ``zero modes''.
Let us change continuously one of the eigenvalues from its value $\lambda=+1$
to $-1$ (Fig.\ref{EVOnTheMove}).  Starting with all eigenvalues on the unit
circle (none of them at the real  positions $\pm 1$)  a complex conjugate pair
moves to the interior. When $\lambda$ changes sign,  the pair becomes a pair of
two real eigenvalues, moving towards the  circle boundary. When $\lambda=-1$, a
new zero mode (and eigenvalue 1 of $Z$) has emerged. 

Note that the eigenvectors in this example have no connection
to space-time physics and there is no association with underlying
physical structures, like in QCD. However the observed behavior of the
eigenvalues is quite characteristic. An exact GW-operator reacts like a step
function. The real eigenvalue  pair disappears and a complex conjugate pair of
eigenvalues appears. For an  approximative GW-operator the eigenvalues flow as
in the example. It is of high interest to quantify the flow and to identify, 
what resolution the respective Dirac operators have with regard to  the
instanton size.

\section{``Artificial'' and realistic instantons}

\begin{figure}[bt]
\begin{center}
\epsfig{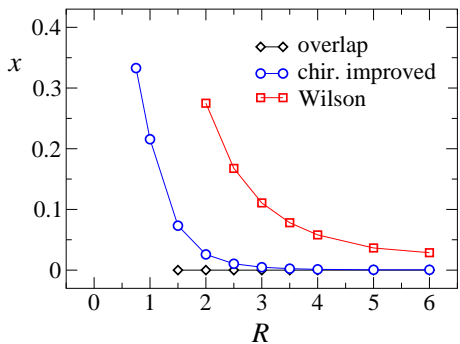}
\end{center}
\caption{\label{SizevsReEV}
Real eigenvalue $x$ associated
with the instanton eigenmode vs. instanton radius in lattice units:  
Wilson operator, overlap operator (with $D_0$ the Wilson operator), 
and the chirally improved operator.}
\end{figure}

We construct gauge configurations with artificial instantons as
discussed in  \cite{FoLaSc85,GaGoLa01} and study for given
lattice size $16^4$ such configurations for instantons of varying size $R$ 
(Fig. \ref{SizevsReEV}). Real eigenvalues may be associated with instanton
modes.

We recover the typical features of our test example: The Wilson operator reacts
quite slowly on the change of size and ``looses'' the instanton at size
$\mathcal{O}(2)$, the overlap operator looses the instanton around $R\simeq
1.5$. The chirally improved  operator is for $R\geq 3$ similar to the overlap
operator, but identifies also smaller instantons. 

The chirally improved operator may be used either for its own sake, as a  Dirac
operator,  or in order to improve performance by using it as a starting point
for the determination of the overlap Dirac operator. From
Fig. \ref{SizevsReEV} one finds, that the improved operator is clearly superior
to  the Wilson operator, since it spends less time (fewer configurations) in
the  central region due to its steeper slope.

\begin{figure}[bt]
\vspace{-1mm}
\begin{center}
\epsfig{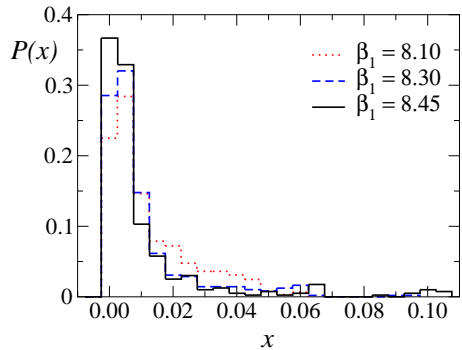}
\end{center}
\caption{\label{ChirImpEVdist}
The distribution of the real eigenvalues $x$ of the chirally improved
operator on $16^4$ lattices (200 configurations for each value of the gauge
coupling $\beta$).}
\end{figure}

In Fig. \ref{ChirImpEVdist} we show the distribution of eigenvalues of the
chirally improved operator. The gauge configurations have been obtained with 
the L\"uscher-Weisz action \cite{LuWe85} as
detailed in \cite{GaGoRa}. Compared to the Wilson action the eigenvalues are
much more localized around 0 and therefore $H$ will have much 
fewer zero modes. This ought to improve the convergence properties 
significantly. However, in our computations we found an improvement factor of
only 2 in the necessary order of the Chebychev polynomials.  This is likely due
to the explicit treatment of the sector of small eigenvalues of $H$. Due to the
numerical  complexity of the improved operator (it has more coupling terms
and is typically a factor of 20 more expensive than Wilson's action), we find
that using it as input for the overlap operator is not economically
warranted. However, to use it as an approximate  Ginsparg-Wilson-type operator
may be quite successful, as we discuss below.

The (normalized)
gauge-invariant densities
\be\label{DefDensity}
p_\alpha(x)=\langle\psi_0(x)|\,\gamma_\alpha \,|\psi_0(x)\rangle
\ee
(implied summation over color and Dirac indices) are determined from the zero
mode eigenvector $|\psi_0\rangle$;  $\gamma_\alpha$ is one of the 16
$\gamma$-matrices of the 4D-Clifford  algebra with $\gamma_0=1$.
Comparing  $p_0$ and $p_5$ as determined for a  realistic gauge
configuration with one zero-mode for the three Dirac operators considered
we find clear instanton bumps at the same position in all cases. 
However, Fig. \ref{SizevsReEV} pointed at a varying sensitivity of different
Dirac operators with regard to identifying instantons. 
The space-time integral $I_0=\int \d^4x\;p_0^2(x)$
defines the so-called inverse participation ratio and constitutes a measure of
localization. For instantons with radius $R$ one expects
$I_0\propto 1/R^4$.

\begin{figure}[bt]
\begin{center}
\epsfig{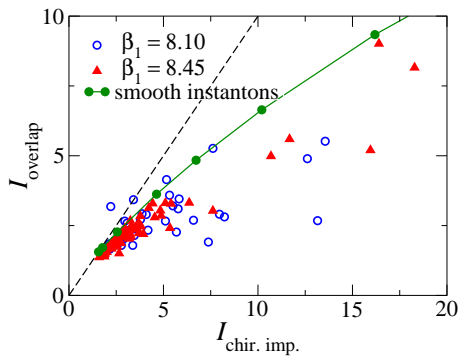}
\end{center}
\caption{\label{IovvsIchirimp}
Comparing the inverse participation ratio for the overlap operator with
the chirally improved operator. For both values of the gauge coupling 
we plot the results of 200 gauge configurations with a single zero mode.}
\end{figure}

In Fig. \ref{IovvsIchirimp} we compare $I_0$ for realistic gauge
configurations.  We find that the overlap operator typically underestimates
the localization of the zero mode, confirming earlier findings for artificial
instantons \cite{GaGoLa01}. A possible interpretation lies in the ``size'' of
the different lattice Dirac operators. The matrix elements of  the overlap
operator fall off like roughly $\sim 1/2^{|x-y|}$ \cite{HeJaLu99} whereas the
chirally improved operator is ultra-local 
(i.e. with non-vanishing coupling only for $|x-y|\leq \sqrt{5}$). 
At typical lattice spacings $\mathcal{O}(0.1\,\textrm{fm})$ of today's
simulations the chirally improved operator appears to keep better track of 
structures smaller than $\mathcal{O}(0.3 \,\textrm{fm})$. Closer to the
continuum limit, at smaller lattice spacings, this  feature of the overlap
operator may be no problem, though.

{\bf Acknowledgment:} We wish to thank our collaborators (M. G\"ockeler, R.
Rakow, S. Schaefer and A. Sch\"afer) for their support and many  discussion 
and the Leibniz Rechenzentrum in Munich for computer time on the Hitachi 
SR8000 and their team for training and support.
C.B.L. wants to thank K. Jansen for instructive discussions.

\vspace{-3mm}

\end{document}